\documentclass{epl}
\usepackage{times}
\usepackage{graphicx}
\usepackage{psfrag} 
\usepackage{fancyhdr}
\begin{document}
\title{Electrostatics of ions inside the nanopores and trans-membrane channels}
\author{\bf Yan Levin} 
\institute{\it Instituto de F\'{\i}sica, Universidade Federal
do Rio Grande do Sul\\ Caixa Postal 15051, CEP 91501-970, 
Porto Alegre, RS, Brazil\\ 
{\small levin@if.ufrgs.br}}
\maketitle
\begin{abstract}

A model of a finite cylindrical ion channel
through a phospholipid membrane of width $L$ 
separating
two electrolyte reservoirs is studied.
Analytical solution of the Poisson equation 
is obtained for an arbitrary distribution of ions inside the 
trans-membrane pore. The solution is asymptotically exact in 
the limit of large ionic strength of electrolyte on the two sides
of membrane.  
However, even for physiological
concentrations of electrolyte, 
the electrostatic barrier sizes found using the theory are in excellent
agreement with the numerical solution of the Poisson equation. 
The analytical solution is used to calculate the electrostatic 
potential energy profiles for pores containing charged protein residues.   
Availability of a semi-exact interionic potential 
should greatly
facilitate the study of ionic transport through nanopores and ion
channels.

\end{abstract}
\bigskip


Ion channels are water filled holes which facilitate exchange
of electrolyte between the exterior and interior of a cell.  
Pores are formed by specific proteins embedded 
into the phospholipid membrane~\cite{Hi01}. 
Depending on the conformation of the
protein, the pore can be open or closed. When open, 
the protein is
very specific to the kind of ions that it allows to pass through
the channel~\cite{DoCaPf98,BeRo01}.  
In order to function properly the channel has
to conduct thousands of ions in a period of few milliseconds. Considering
that the channel passes through a phospholipid membrane which has
a very low dielectric constant and is very narrow, producing   
high energetic penalties for ions entering the nonopores, 
it is fascinating
to contemplate how Natures manages to perform this amazing task.
In fact, as long ago as 1969, Parsegian observed that for an 
infinitely long cylindrical channel~\cite{Sm50} of radius $a=3$ \AA,
the electrostatic barrier is over $16k_B T$, which should completely
suppress any ionic flow~\cite{Pa69}.  
Later numerical work by Levitt~\cite{Le78}, Jordan~\cite{Jo82}
and others demonstrated that for more realistic finite channels 
the barrier
is dramatically reduced. For example, for a channel of length $L=25$ \AA 
$\,$ and radius $a=3$ \AA, the barrier is about $6k_BT$,  which although
still quite large, should allow ionic conductivity.
Recently the study  of ion channels
has expanded to other parts of applied physics.  Water filled
nanopores are introduced into silicon oxide films, polymer membranes, 
etc~\cite{LiSt01,SiFu02}. In
all of these cases the dielectric constant of the interior of a nanopore
greatly exceeds that of the surrounding media.

To quantitatively 
study the conductance of a nanopore one has three options: the all atom
molecular dynamics simulation (MD)~\cite{BeRo01}; 
the Brownian dynamics simulation (BD)~\cite{MoCo00,CoKu00} with implicit 
water treated as a uniform dielectric continuum;
or the mean-field Poisson-Nernst-Planck theory (PNP)~\cite{Ei99} 
which treats both water and ions implicitly. 
While clearly the most accurate, MD simulations are computationally very
expensive~\cite{Lev99}.
Brownian dynamics is significantly
faster than MD, but because of the dielectric discontinuities across the
various interfaces a new solution of the Poisson equation
is required for each new configuration of ions inside the pore.   
The simplest approach to study the ionic conduction
is based on the PNP theory~\cite{Ei99}. This combines the continuity
equation with the Poisson equation and Ohm's and  Fick's laws. 
PNP is intrinsically mean-field and is, therefore, bound to fail when 
ionic correlations become important.  This has been well studied for its
static version --- the Poisson-Boltzmann equation, which
is known to break down for aqueous electrolytes with multivalent
ions and also for monovalent electrolytes in 
low dielectric solvents~\cite{Le02,NaJu05}.  
For narrow channels, the cylindrical geometry, combined with the 
field confinement, results in a pseudo one dimensional potential of
very long range~\cite{Te05,ZhKaSh05}.
Under these conditions the correlational effects dominate, and
the mean-field approximation fails~\cite{Le02}. 
Indeed recent comparison between the
BD and the PNP showed that PNP breaks down when the pore radius is smaller
than about two Debye lengths~\cite{MoCo00,CoKu00}.  
At the moment, therefore, it appears that 
a semi-continuum (implicit solvent) Brownian dynamics simulation is
the best compromise between the cost and 
accuracy~\cite{Lev99,KuAnCh01,TiBiSm01} for narrow pores. 
Unfortunately
even this, simplified strategy demands a tremendous computational effort.
The difficulty is that BD requires a new solution of 
the Poisson partial differential equation 
at each time step.   This can be partially overcome
by using lookup tables~\cite{CoKu00} and variational methods~\cite{AlMeHa02},
but still requires a supercomputer.
If the interaction potential
between the ions inside the channel would be known, the simulation 
could proceed orders of magnitude faster.  
However, up to now the only exact 
solution to the Poisson equation in
a cylindrical geometry was
for the case of an infinitely long pore~\cite{Sm50,Pa69,Te05}.  
In this letter we shall
provide another exact solution, but now 
for a finite trans-membrane channel.

We shall work in the context of a 
primitive model of electrolyte and membrane.  The membrane will be
modeled as a uniform dielectric slab of width $L$ located between $z=0$
and $z=L$. The dielectric constant of the membrane and the channel
forming protein is taken to be $\epsilon_p \approx 2$.  On both sides of the 
membrane there is an  electrolyte solution composed of point-like ions and
characterized by the inverse Debye length $\kappa$. 
A channel is a cylindrical hole of radius $a$ and length $L$ 
filled with water. As is usual for continuum electrostatic 
models~\cite{Lev99},
we shall take the dielectric constant of water 
inside and outside the channel
to be the same, $\epsilon_w \approx 80$. 
It is convenient to set up a cylindrical
coordinate system $(z,\rho,\phi)$ with the origin located at the center
of the channel at $z=0$.  Suppose that an ion is located at an
arbitrary position ${\bf x'}$ inside the channel.  The electrostatic potential
$\varphi(z,\rho, \phi;{\bf x'})$ inside the channel and  membrane 
satisfies the Laplace equation
\begin{equation}
\label{1}
\nabla^2 \varphi=-\frac{4\pi q}{\epsilon_w} \delta({\bf x}-{\bf x'})\;.
\end{equation}
For $z>L$ and $z<0$, $\varphi({\bf x};{\bf x'})$ satisfies the linearized
Poisson-Boltzmann or the Debye-H\"uckel equation~\cite{Le02}
\begin{equation}
\label{2}
\nabla^2 \varphi=\kappa^2 \varphi\;.
\end{equation}
The inverse Debye length is related to the ionic strength $I$ of
electrolyte, $\xi_D^{-1}=\kappa= \sqrt{8 \pi \lambda_B I}$,
where $\lambda_B=q^2/\epsilon k_B T$ is the Bjerrum length and
$I=(\alpha^2 c_\alpha+\alpha c_\alpha +2 c)/2$. Here $c_\alpha$ is the
concentration of $\alpha:1$ valent electrolyte and $c$ is the concentration
of $1:1$ electrolyte. All the usual boundary conditions must be enforced:
the potential must vanish at infinity and be continuous 
across all the interfaces;  the tangential component of the electric field
and the normal component of the electric displacement must be  
continuous across all
the interfaces.  These boundary conditions guaranty the uniqueness of
solution.  Unfortunately, even this relatively simple geometry can not,
in general, be solved exactly.  We observe, however, that an {\it exact}
solution is possible in the limit that $\kappa \rightarrow \infty$.  
In this special case the system of differential 
equations becomes separable.  Our, strategy then will be to solve exactly 
this asymptomatic problem and then extend the solution to finite values of 
Debye length.

We start by making the following fundamental observation. The condition  
$\kappa \rightarrow \infty$ signifies that electrolyte perfectly screens
any electric field --- the Debye length is zero.  This, combined with the
boundary condition --- electrostatic potential 
must vanishes at infinity --- implies that in this limit
$\varphi(0,\rho,\phi;{\bf x'})=\varphi(L,\rho,\phi;{\bf x'})=0$, 
for {\it any} position ${\bf x'}$ of an ion inside the pore. This is 
a dramatic simplification.  Now it is no longer necessary to solve 
the Debye-H\"uckel equation, but only the
Poisson equation with a perfect grounded conductor boundary 
conditions at $z=0$ and 
$z=L$. To proceed
we expand the $\delta(z-z')$ in eigenfunctions of the differential
operator
\begin{equation}
\label{3}
\frac{\rm d^2 \psi_n}{\rm d z^2}+k_n^2 \psi_n=0\;,
\end{equation}
satisfying the perfect conductor boundary condition.  
The normalized eigenfunctions are 
$\psi_n(z)=\sqrt{2/L}\sin(k_n z)$, with 
$k_n=n \pi/L$.
The Sturm-Liouville nature of the
differential Eq. (\ref{3}) guaranties us that
\begin{equation}
\label{4}
\delta(z-z')=\frac{2}{L}\sum_{n=1}^\infty \sin(k_n z)\sin(k_n z') \;.
\end{equation}
Similarly,
\begin{equation}
\label{5}
\delta(\phi-\phi')=
\frac{1}{2\pi}\sum_{m=-\infty}^\infty e^{i m(\phi-\phi')} \;.
\end{equation}
Next we write
\begin{equation}
\label{6}
\varphi({\bf x},{\bf {x'}})=\frac{q}{\pi\epsilon_w L}\sum_{n=1}^\infty\sum_{m=-\infty}^\infty e^{i m(\phi-\phi')} \sin(k_n z)\sin(k_n z') g_{nm}(\rho,\rho')\;.
\end{equation}
Substituting this into Eq.(\ref{1}) we find that the Green function 
$ g_{nm}(\rho,\rho')$
satisfies the modified Bessel equation 
\begin{equation}
\label{7}
\frac{1}{\rho}\frac{{\rm d}}{{\rm d} \rho}\rho \frac{{\rm d} g_{nm}}{{\rm d} \rho }- (k_n^2 +\frac{m^2}{\rho^2}) g_{nm}=
-\frac{4 \pi}{\rho}\delta(\rho-\rho') \,
\end{equation}
solution of which can be found using the usual techniques~\cite{Sm50,Ja99}.  
We obtain,
\begin{equation}
\label{8}
g_{mn}(\rho,\rho')=4\pi I_m(k_n \rho_<)[K_m(k_n \rho_>)+\gamma_{mn}I_m(k_n \rho_>)]\,
\end{equation}
where $\rho_>$ and $\rho_<$ are the larger and the smaller of the
set $(\rho,\rho')$ and 
\begin{equation}
\label{9}
\gamma_{mn}=\frac{K_m(k_n a)K_m'(k_n a)(\epsilon_p-\epsilon_w)}
{\epsilon_w I_m'(k_n a)K_m(k_n a)-\epsilon_p I_m(k_n a)K_m'(k_n a)}\;.
\end{equation}
Here $I_m,K_m,I_m',K_m'$ are the modified Bessel functions of the first and
second kind and their derivatives, respectively.  Eqn.(\ref{8}) is valid
for $\rho_>\le a$. When $\rho> a$,
\begin{equation}
\label{10}
g_{mn}(\rho,\rho')=\frac{4 \pi \epsilon_w}{k_n a}\frac{ K_m(k_n \rho)I_m(k_n \rho')}{\epsilon_w I_m'(k_n a)K_m(k_n a)-\epsilon_p I_m(k_n a)K_m'(k_n a)} \;.
\end{equation}
Eqns.(\ref{6},\ref{8}), and (\ref{10}) 
are exact for an ion
inside a pore with perfect conductor boundary 
conditions at $z=0$ and $z=L$. If the ion is located on the axis of
symmetry, $z'=z_0$, $\rho'=0$, only $m=0$ term in Eqn.(\ref{6})
survives, and the electrostatic potential {\it inside} the channel at position 
$z,\rho$ takes a particularly simple form,
$\varphi_{in}(z,\rho;z_0)=\varphi_{1}(z,\rho;z_0)+\varphi_2(z,\rho;z_0)$, where
\begin{equation}
\label{11}
\varphi_1(z,\rho;z_0)=\frac{4 q}{\epsilon_w L}\sum_{n=1}^\infty
\sin(k_n z)\sin(k_n z_0)K_0(k_n \rho)\;,
\end{equation}
and 
\begin{equation}
\label{12}
\varphi_2(z,\rho;z_0)=\frac{4 q(\epsilon_w-\epsilon_p)}{\epsilon_w L}\sum_{n=1}^\infty
\frac{K_0(k_n a)K_1(k_n a)I_0(k_n \rho)\sin(k_n z)\sin(k_n z_0)}
{\epsilon_w I_1(k_n a)K_0(k_n a)+\epsilon_p I_0(k_n a)K_1(k_n a)}\;.
\end{equation}
Eqs. (\ref{11},\ref{12}) 
are exact in the $\kappa \rightarrow \infty$ limit. 
To see how
these equations can be 
extended to finite values of $\kappa$, it is 
important to first understand
their physical meaning.
Potential $\varphi_2$ is mostly the result of the charge induced on the
interface between the high dielectric aqueous interior of the pore 
and the low dielectric membrane. We expect that this term will be affected
very little by the precise value of the Debye length of the
surrounding electrolyte solution. 

The potential
$\varphi_1$ contains the contribution from the ion located at $z_0$ and from
the induced charge on the pore/electrolyte and the membrane/electrolyte
interfaces. It will, therefore, strongly depend on the
precise value of $\kappa$. Furthermore, we observe 
that Eq.~(\ref{11}) is  {\it exactly}
the potential produced by a charge $q$ located inside an {\it infinite} 
slab of water of width $L$ bounded by two grounded 
perfectly conducting planes~\cite{Ja99}. 
This key observation allows us to explicitly
resum the series in Eq.~(\ref{11}).  However, it is possible to  
do even better, and now enforce the {\it exact} boundary condition, 
namely that for $z<0$ and $z>L$ the
electrostatic potential must satisfy the  
Debye-H\"uckel equation (\ref{2}). Using the Bessel $J$ representation
of the delta function one can constructs the Green function~\cite{LeMe01} 
which satisfies all the boundary conditions for the slab geometry 
and has the required symmetry property~\cite{Ja99} between 
the source and the observation points. We then find 
\begin{equation}
\label{13}
\varphi_1(z,\rho;z_0)=\int_0^\infty {\rm d}k \frac{J_0(k \rho)\left\{\alpha^2(k) e^{k|z-z_0|-2 k L}+
\alpha(k)\beta(k)[ e^{-k(z+z_0)}+e^{k(z+z_0)-2 k L}]+ \beta^2(k) e^{-k|z-z_0|}\right\}}
{\beta^2(k)-\alpha^2(k)\exp(- 2 k L)} \;,
\end{equation}
where $\alpha(k)=[k-\sqrt{k^2+\kappa^2}]/2 k$, 
$\beta(k)=[k+\sqrt{k^2+\kappa^2}]/2 k$, and $J_0(x)$ is the Bessel 
function of first kind and order zero.  
Eq.~(\ref{13}) provides an analytic continuation of Eq.~(\ref{11}) into
finite $\kappa$ parameter space.  It can be  checked explicitly that in  
the limit $\kappa \rightarrow \infty$, Eq.~(\ref{13}) exactly sums
the series in Eq.~(\ref{11}).  
Finally, for the region $\rho > a$ the electrostatic potential is
\begin{equation}
\label{14}
\varphi_{out}(z,\rho;z_0)=
\frac{4 q}{L}\sum_{n=1}^\infty \frac{1}{k_n a}
\frac{K_0(k_n \rho)\sin(k_n z)\sin(k_n z_0)}
{\epsilon_w I_1(k_n a)K_0(k_n a)+\epsilon_p I_0(k_n a)K_1(k_n a)}\;.
\end{equation}
Eq~(\ref{14}) is 
exact only for the perfect conductor boundary conditions, however, 
the huge jump
in the  dielectric constant going from the membrane's interior 
to the aqueous 
electrolyte will leave $\varphi_{out}$ mostly unaffected even for
finite values of $\kappa$. 

If the channel contains $N$ ions and charged protein residues 
their interaction energy
is given by 
\begin{equation}
\label{15}
V=\frac{1}{2}\sum_{i,j}^N q_i \varphi^j\;,
\end{equation}
where $q_i$ is the charge of ion/residue $i$ and $\varphi^j$ is the
electrostatic 
potential produced by the ion/residue $j$ at the position of ion/residue $i$.
Similarly the electrostatic barrier that an ion feels as
it moves through a charge free
channel is~\cite{Le02},
\begin{equation}
\label{16}
U(z)=\frac{q}{2}\lim_{\rho \rightarrow 0}\left[\varphi(z,\rho;z)-\frac{q}{\epsilon_w \rho}\right]+\frac{q \kappa}{2 \epsilon_w}\;.
\end{equation}

The last term in Eq.~(\ref{16}) is the electrostatic 
``solvation'' energy  that
a point-like ion looses as it moves from the bulk electrolyte 
into the interior or a pore.  This energy
can be calculated using the Debye-H\"uckel
theory and is equivalent to the 
excess chemical potential resulting from the screening of ionic 
electric field by the surrounding electrolyte~\cite{Le02}. 
The limit in Eq.~(\ref{16}) is easily obtained by noting that 
\begin{equation}
\label{17}
\frac{1}{\rho}=\int_0^\infty J_0(k \rho) {\rm d} k\;.
\end{equation}

\begin{figure}[th]
\begin{center}
\psfrag{U}{$U(z)\; (k_B T)$}
\psfrag{U1}{\hspace{-5mm}$U(z)\; (kJ/mol)$}
\psfrag{z}{$z$}
\twofigures[width=5cm,angle=270]{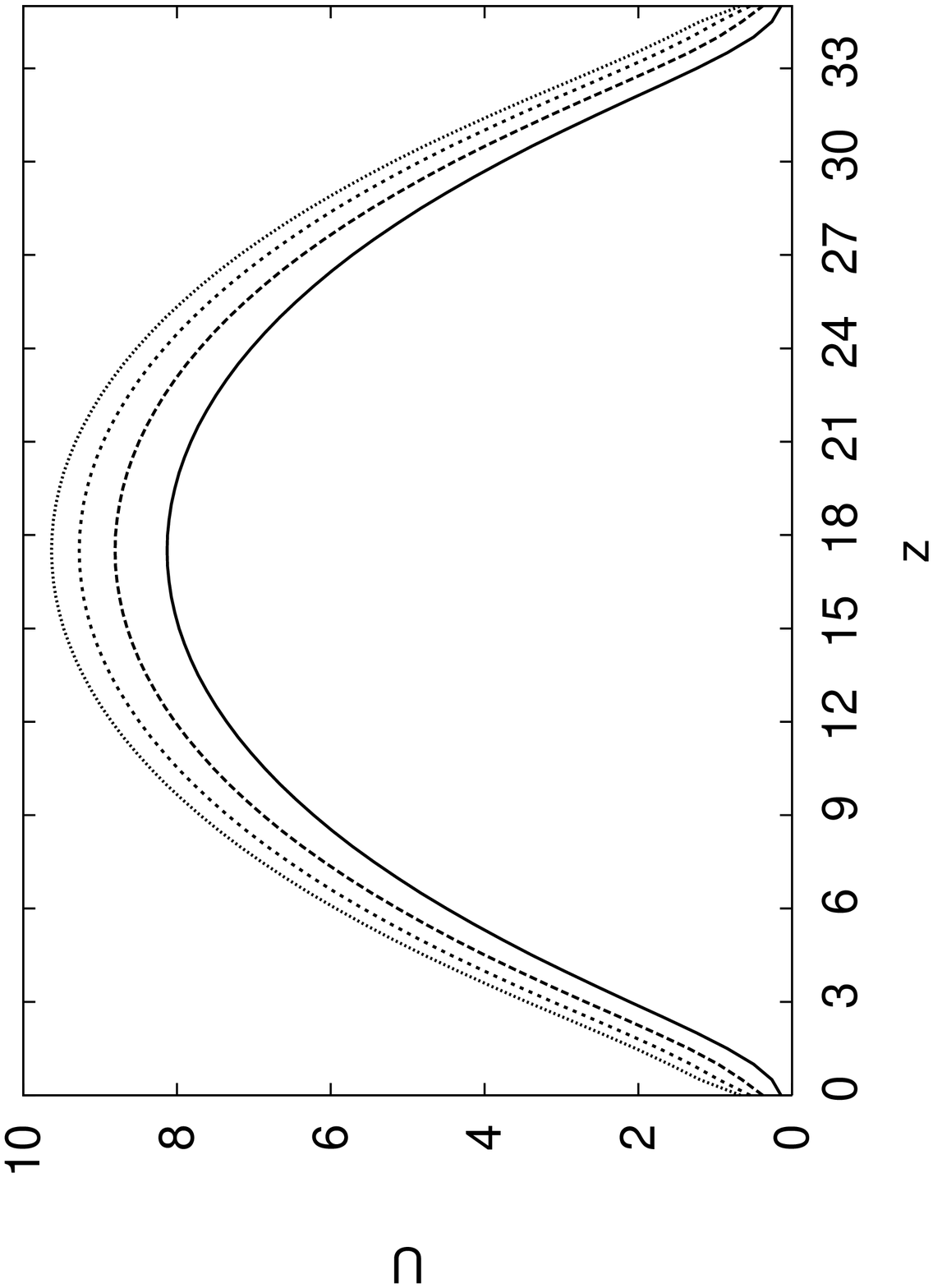}{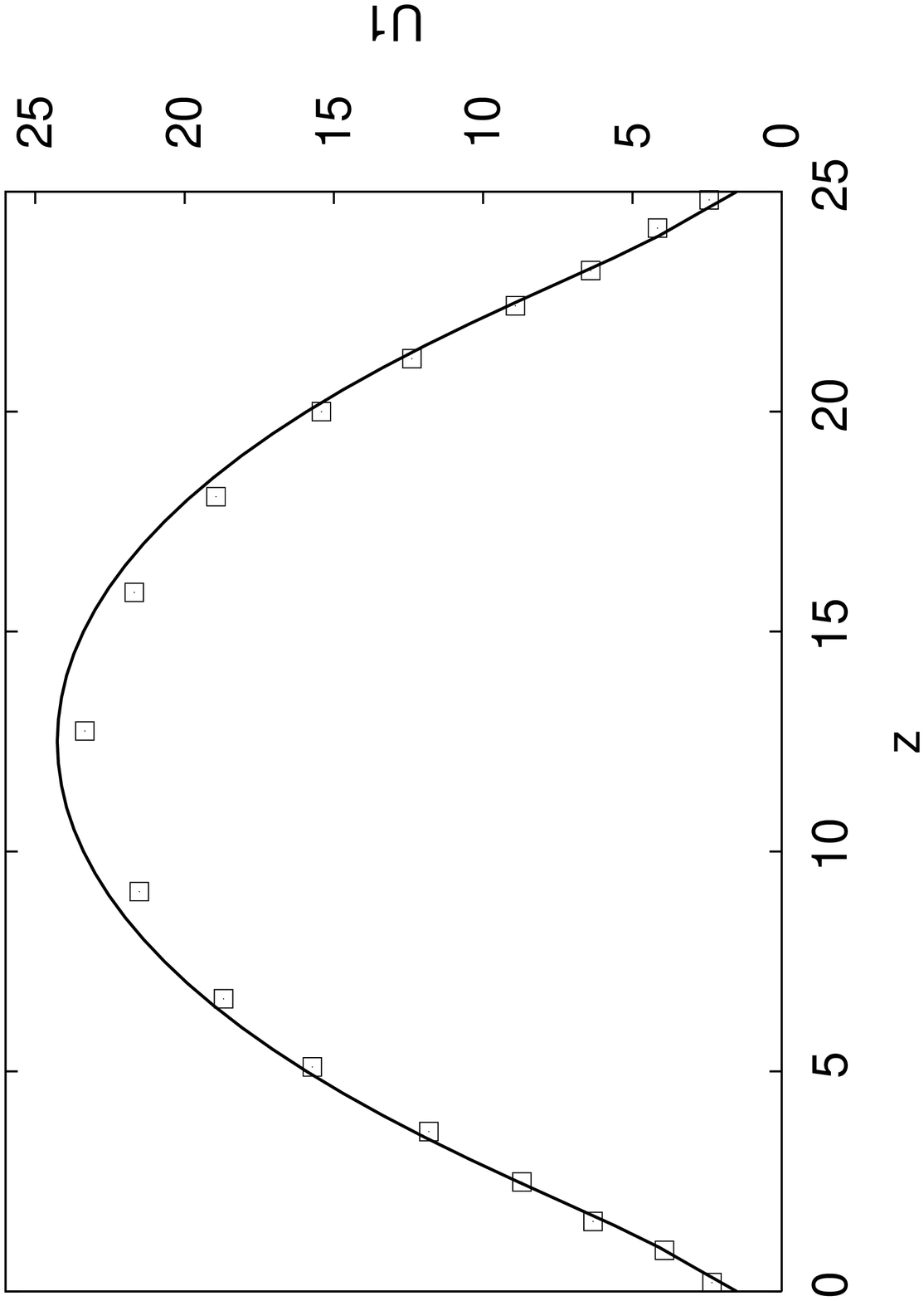}  
\caption{Electrostatic potential barriers for an ion of charge $q$ moving
along the axis of symmetry through a  channel of $L=35$ \AA $\,$
and $a=3$ \AA $\,$.  External electrolyte
concentration from the bottom up is $0.15$, $1$, $2$ and $3$M.\label{fig1}}
\caption{Electrostatic potential barrier for an ion of charge $q$ moving
through a pore of  $L=25$ \AA $\,$
and $a=2.5$ \AA $\,$.  External electrolyte
concentration is $2.5$M. Symbols are the result of a numerical integration
of the system of Poisson-Boltzmann (bulk electrolyte) 
and Poisson equation (pore interior) from ref.~\cite{JoBa89}\label{fig1a}}
\end{center}
\end{figure}

We are now in a position to explore some of the quantitative 
consequences of
the current theory.  
In Fig.~(\ref{fig1}) 
we first plot the potential energy barrier for an ion of charge $q$ moving
through a channel
of $L=35$ \AA $\,$ and  $a=3$ \AA $\,$ and various
 external electrolyte concentrations.
For the physiological salt concentration ($150$mM) we find the barrier 
height to be $8.13 k_BT$.  Using
numerical solution of the Poisson equation, Levitt obtained a barrier
of $8.48 k_BT$.  Some of the difference between the two values
can be attributed to the fact that 
in numerical calculations  presence of external electrolyte 
was not taken into account.  As the length of the channel increases,
the role of external electrolyte becomes 
relatively less important.  Indeed for 
a channel of  $L=50$ \AA $\,$ 
and  $a=2$ \AA, we obtain a barrier of $18.65 k_B T$, while 
the Levitt's numerical solution produced $17.2k_BT$~\cite{Le78} and 
Jordan's  $18.6 k_B T$~\cite{Jo82}.
The only numerical work known to us that 
explicitly takes into presence existence of external
electrolyte is ref.~\cite{JoBa89}.  The authors of that paper 
numerically solved the non-linear Poisson-Boltzmann 
equation for external electrolyte and the Poisson equation for the
interior of the channel.  For a pore of $L=25$ and $a=2.5$ \AA, 
and electrolyte concentration of $2.5$M, they find a 
barrier of $9.5k_B T$, while we obtain $9.8 k_B T$. 
In Fig.~(\ref{fig1a}) we compare the full electrostatic energy 
barrier obtained from the numerical solution with our analytical 
results.  The agreement, once again, is quite good.

\begin{figure}[th]
\begin{center}
\psfrag{U}{$\Phi(z)\; (k_B T)$}
\psfrag{z}{$z$}
\twofigures[width=5cm,angle=270]{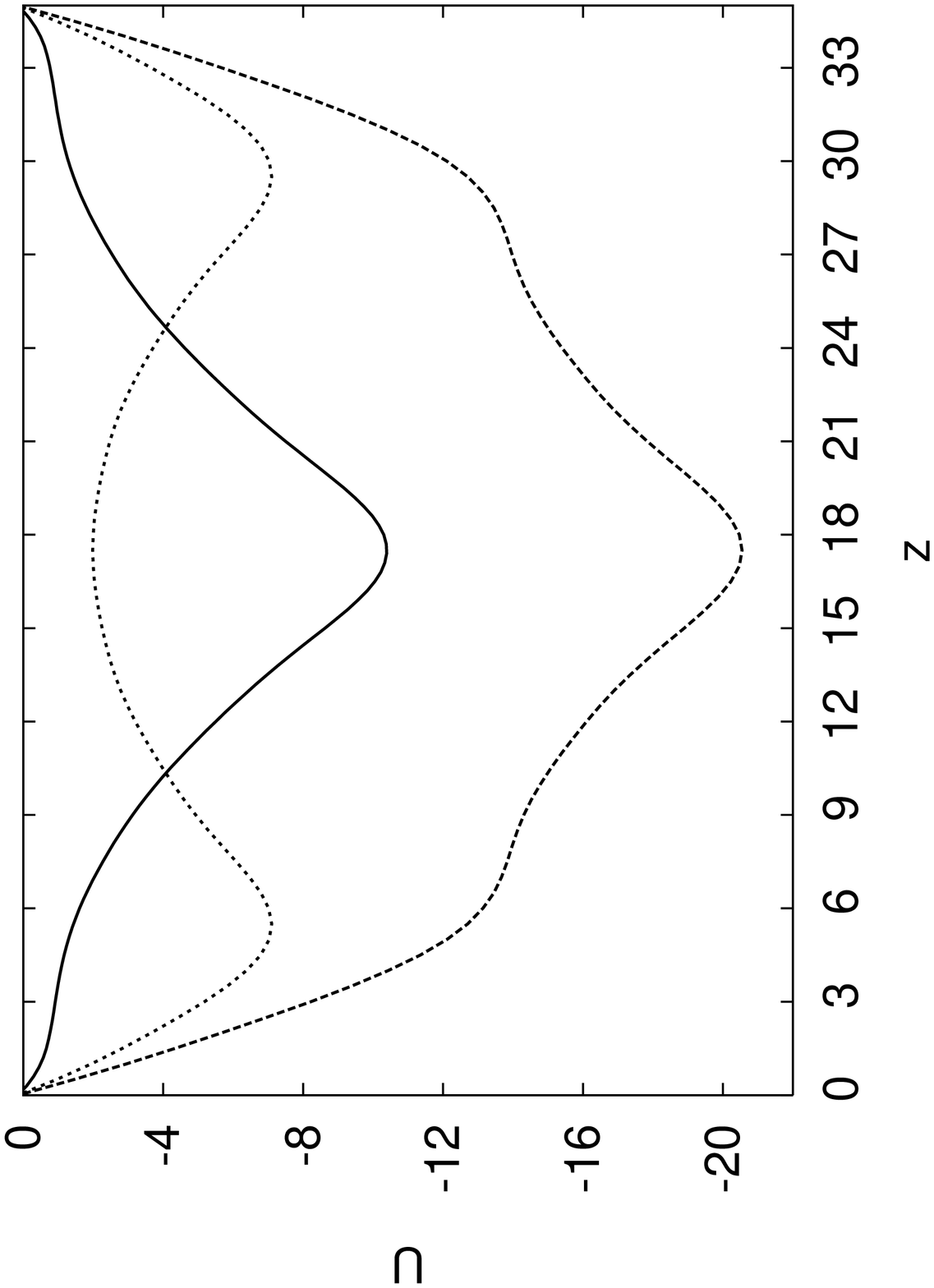}{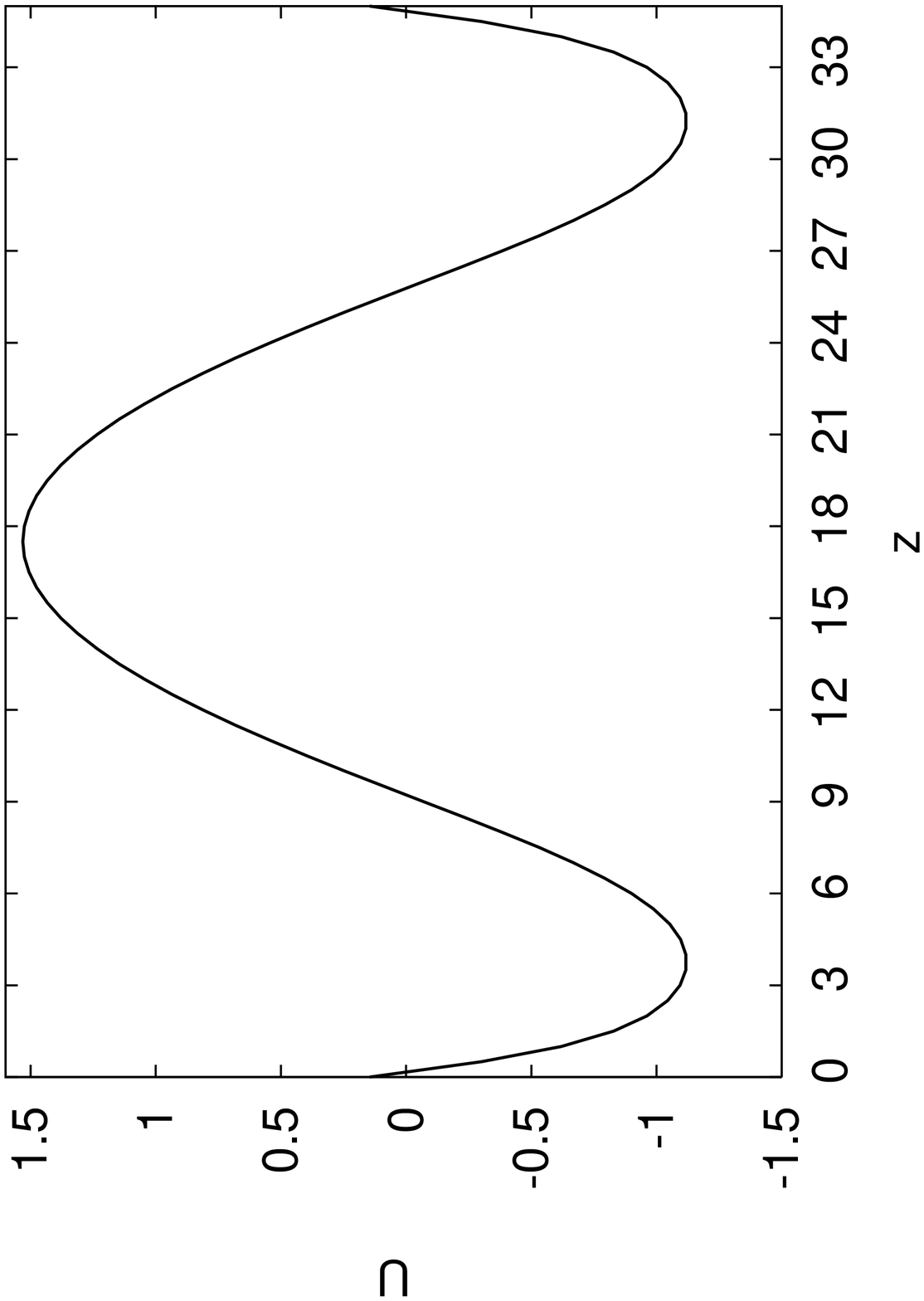} 
\caption{Potential energy profile for an ion of charge $q$ moving
along the axis of symmetry through a channel of $L=35$ \AA $\,$
and $a=3$ \AA $\,$ containing a protein residue of charge $-q$ embedded 
into its wall $(\rho=a)$ at $z=L/2$ (solid line); 
containing two charged 
residues  at 
$z=5$ and z=$30$ \AA $\,$ (dotted line);  containing 
three residues at  $z=5,17.5$ and $z=30$ \AA $\,$ (dashed line).
External electrolyte
concentration is $150$mM.\label{fig2}}
\caption{Potential energy profile for an ion of charge $q$ moving along the
axis of symmetry through a pore
of $L=35$ \AA $\,$ and $a=3$ \AA, 
containing two charged residues 
hidden in the membrane's interior ($\rho=6$ \AA) located at
$z=5$ and $z=30$ \AA. External electrolyte
concentration $150$mM.\label{fig3}}
\end{center}
\end{figure}

Availability of a semi-exact interaction potential allows us to  
easily explore
the potential energy landscape $\Phi=V+U$ of an ion of charge $q$ 
moving 
through a channel which also
contains some fixed charged protein residues.  
For example, consider a channel
of  $L=35$ \AA $\,$ and $a=3$ \AA $\,$  and suppose that 
there is one protein residues of charge $-q$, 
embedded into the surface of the channel at $(z=L/2,\rho=a, \phi)$. 
In Fig.~\ref{fig2} we show the potential energy profile for an ion
moving along the axis of symmetry through such a channel.  
Instead of
a potential barrier, this ion  encounters a potential well
of depth more than $10 k_B T$!  It will, therefore, find it extremely
difficult to pass through such a channel.

Now, suppose that two  charged residues are 
embedded into the channel wall, 
one close to the  entrance of the channel at
$z=5$ $\,$ \AA $\,$ and another close to its exit at $z=30$ \AA, 
both at $\rho=3$ \AA. 
The electrostatic potential, now develops a double-well structure, 
Fig.~\ref{fig2}. 
Each minimum is relatively less deep than in a channel with only one
central residue.  One might then suppose that 
adding more charged residues will diminish the depth of the wells even
further.  This, however, is not the case.  In Fig.\ref{fig2} we also show
the potential energy landscape of a channel containing three 
uniformly spaced residues. 
Evidently instead of decreasing the depth of the potential well, it 
has dramatically increased! 
There is, however, a mechanism which Nature can use to
diminish the depth of potential wells --- hide the charged residues
in the membrane's (or protein's) hydrophobic interior~\cite{KuBa03}. In 
Fig.~\ref{fig3} we plot the potential energy profile for the same channel
as in Fig.~\ref{fig2}, but with the two charged residues hidden 
in the  membrane's hydrophobic interior at $\rho=6$ \AA.
In this case the deep potential well is replaced by a shallow binding
site followed by an activation barrier of only $2.5 k_B T$. This 
can be easily overcome by an external electric field or
a chemical potential gradient. We find that there is  
an optimum location 
for hiding charged residues in order to 
produce the smallest barrier for channel penetration. The formalism
developed above allows us to easily explore all the parameter
space in order to find this optimum position. 

To conclude, we have presented an analytically solvable model
of electrostatics inside an ion channel.   
The solution found is exact in the limit of 
large electrolyte concentrations.  However, comparison 
with the numerical
work shows that it remains valid
even at intermediate and low electrolyte concentrations. 
The analytical solution can be used to dramatically speed up
the Brownian dynamics simulations of ionic transport through
cylindrical pores.  The biological and structural
information can be partially taken into account 
through a proper placement of charged protein residues. 
Furthermore, even if a more detailed
atomistic molecular dynamics simulation is necessary, availability
of a rapid Brownian dynamics model can serve for a initial exploration
of the parameter space. 

In this work water inside the pore has been treated as a uniform dielectric
continuum identical to the bulk.  
While this is acceptably for wider pores, the approximation
will certainly fail for very narrow pores such as gramicidin.  
To properly account for the polarization of water in this geometry 
one must go beyond the continuum dielectric approximation~\cite{PeHu05}. 
Until now, the only option for these cases was to perform all atom
molecular dynamics simulations.  
The current work suggests that another way might be possible.  
The continuum description with $\epsilon_w=80$ and $\epsilon_p=2$ 
can be used  for the bulk water outside the
channel, for the membrane, and for the trans-membrane protein, 
while inside the channel (now  with $\epsilon_{in}=1$) 
one might try to obtain an accurate analytical electrostatic potential.  
The Coulomb interactions between
the water molecules inside the channel could then be treated 
explicitly, without any
need for continuum dielectric approximation. This would then allow
to perform very fast molecular dynamics simulations of ionic transport,
free of the drawbacks associated with the implicit solvent models.
The work in this direction is now in progress.

This work was supported in part by
Conselho Nacional de
Desenvolvimento Cient{\'\i}fico e Tecnol{\'o}gico (CNPq).

\bibliographystyle{prsty}
\bibliography{references}

\begin{thebibliography}{10}

\bibitem{Hi01}
B. Hille, {\em Ionic {C}hannels of {E}xcitable {M}embranes, 3rd ed.} (Sinauer
  Associates, Sunderland, Ma, 2001).

\bibitem{DoCaPf98}
D.~A. Doyle and {et. al}, Nature {\bf 280},  69  (1998).

\bibitem{BeRo01}
B. Roux and S. Berneche, Nature {\bf 414},  73  (2001).

\bibitem{Sm50}
W.~R. Smythe, {\em Static and {D}ynamic {E}lectricity} (McGraw-Hill, New York,
  1950).

\bibitem{Pa69}
A. Parsegian, Nature {\bf 221},  844  (1969).

\bibitem{Le78}
D.~G. Levitt, Biophys. J. {\bf 22},  209  (1978).

\bibitem{Jo82}
P.~C. Jordan, Biophys. J. {\bf 39},  157  (1982).

\bibitem{LiSt01}
J. Li and {et al.}, Nature {\bf 412},  166  (2001).

\bibitem{SiFu02}
Z. Siwy and A. Fulinski, Phys. Rev. Lett. {\bf 89},  Art. No. 198103  (2002).

\bibitem{MoCo00}
G. Moy, B. Corry, S. Kuyucak, and S.~H. Chung, Biophys. J. {\bf 78},  2349
  (2000).

\bibitem{CoKu00}
B. Corry, S. Kuyucak, and S.~H. Chung, Biophys. J. {\bf 78},  2364  (2000).

\bibitem{Ei99}
R.~S. Eisenberg, J. Membr. Biol. {\bf 171},  1  (1999).

\bibitem{Lev99}
D.~G. Levitt, J. Gen. Physiol. {\bf 113},  789  (1999).

\bibitem{Le02}
Y. Levin, Rep. Prog. Phys. {\bf 65},  1577  (2002).

\bibitem{NaJu05}
A. Naji, S. Jungblut, A.~G. Moreira, and R.~R. Netz, Physica A {\bf 352},  131
  (2005).

\bibitem{Te05}
S. Teber, J. Stat. Mec. Theory and Experiment  Art. No. P07001  (2005).

\bibitem{ZhKaSh05}
J. Zhang, A. Kamenev, and B.~I. Shklovksii, Phys. Rev. Lett. {\bf 95},  Art.
  No. 148101  (2005).

\bibitem{KuAnCh01}
S. Kuyucak, O.~S. Andersen, and S.~H. Chung, Rep. Prog. Phys. {\bf 64},  1427
  (2001).

\bibitem{TiBiSm01}
D.~P. Tieleman, P.~C. Biggin, G.~R. Smith, and M.~S.~P. Sansom, Quaterly Rev.
  Biophysics {\bf 34},  473  (2001).

\bibitem{AlMeHa02}
R. Allen, S. Melchionna, and J.~P. Hansen, Phys. Rev. Lett.  Art. no. 175502
  (2002).

\bibitem{Ja99}
J.~D. Jackson, {\em Classical {E}lectrodynamics} (Wiley, New York, 1999).

\bibitem{LeMe01}
Y. Levin and J.~E. Flores-Mena, Europhys. Lett. {\bf 56},  187  (2001).

\bibitem{JoBa89}
P.~C. Jordan, R.~J. Bacquet, J.~A. McCammon, and P. Tran, Biophys. J. {\bf 55},
   1041  (1989).

\bibitem{KuBa03}
S. Kuyucak and T. Bastug, J. Biological Physics {\bf 29},  429  (2003).

\bibitem{PeHu05}
C. Peter and G. Hummer, Biophys. J. {\bf 89},  2222  (2005).

\end{thebibliography}
\end{document}